\documentclass[3p,times,twocolumn]{elsarticle}
\usepackage{natbib}
\usepackage{ecrc}
\volume{00}
\firstpage{1}
\journalname{Nuclear and Particle Physics Proceedings}
\runauth{R. Aloisio}
\jid{nuphbp}
\jnltitlelogo{Nuclear and Particle Physics Proceedings}
\usepackage{amssymb}
\usepackage[figuresright]{rotating}

\begin{document}
\begin{frontmatter}

\title{Ultra High Energy Cosmic Rays and Neutrinos}
\author{Roberto Aloisio}
\address{Gran Sasso Science Insitute (INFN), viale F. Crispi 7, 67100 L'Aquila, Italy\\
INAF/Osservatorio Astrofisico di Arcetri, largo E. Fermi 5, 50125 Firenze, Italy}

\begin{abstract}
We discuss the production of ultra high energy neutrinos coming from the propagation of ultra high energy cosmic rays and in the framework of top-down models for the production of these extremely energetic particles. We show the importance of the detection of ultra high energy neutrinos that can be a fundamental diagnostic tool to solve the discrepancy in the observed chemical composition of ultra high energy cosmic rays and, at the extreme energies, can unveil new physics in connection with the recent cosmological observations of the possible presence of tensor modes in the fluctuation pattern of the cosmic microwave background. 
\end{abstract}

\begin{keyword}
Ultra High Energy Cosmic Rays \sep Ultra High Energy Neutrinos \sep Super Heavy Dark Matter 
\end{keyword}

\end{frontmatter}

\section{Introduction}
\label{Intro}

Ultra High energy Cosmic Rays (UHECR) are the most energetic particles observed in nature, with energies up to $10^{20}$ eV. The experimental and theoretical study of UHECR have brought several important results, such as \cite{Kotera:2011cp,Kampert:2014eja,Blasi:2014roa}: (i) UHECRs are charged particles, with limits on neutral particles up to $10^{19}$~eV at the level of few percent for photons and well below for neutrinos \cite{Abraham:2009qb,Abu-Zayyad:2013dii,Abreu:2013zbq}, (ii) the spectra observed on Earth show a slight flattening at energies around $5\times 10^{18}$~eV (called the ankle), with (iii) a steep suppression at the highest energies around $10^{20}$~eV. 

One of the key informations in the physics of cosmic rays in general and UHECR in particular is the composition of this radiation. In this respect, experimental observations are still not conclusive. The Pierre Auger Observatory (Auger) \cite{Aab:2015zoa}, far the largest detector devoted to the observation of UHECR, points toward a mixed composition with light (proton and He) elements dominating the low energy tail of the spectra and a heavier composition at the highest energies, that starts around energies $5\times 10^{18}$ eV. On the other hand, Telescope Array (TA) \cite{AbuZayyad:2012kk}, even if with 1/10 of the Auger statistics, claims a proton dominated composition at all energies up to the highest observed. 

The actual chemical composition of UHECR is a key information in order to understand the physical mechanisms responsible for the acceleration of this particles and, ultimately, to tag their sources. The production of secondary particles, such as neutrinos or gamma rays, is due to the interaction of UHECR with astrophysical backgrounds during propagation and, being strongly tied with UHECR chemical composition, can be of paramount importance to solve the alleged contradiction between Auger and TA observations. In this paper we will review secondary neutrino production bracketing the expectations connected with different assumption on chemical composition. 

On more general grounds, the fraction of neutrinos (and gamma-rays) observed in the spectra of UHECR at energies till $10^{19}$ eV, as stated above, is extremely small with only upper limits and no direct observations. Nonetheless, the observations of Auger and TA at the highest energies are still affected by a reduced number of events. For instance, Auger, with the highest statistics, has only 5 events at energies around $10^{20}$ eV \cite{Aab:2015bza}. This energy range is of paramount importance in order to unveil possible top-down mechanism in the production of UHECR connected with new physics at the inflation scale. 

The recent claim by BICEP2 of a substantial contribution of tensor modes to the fluctuation pattern of the Cosmic Microwave Background, even if reconsidered after the combined analysis with Planck and Keck array \cite{Ade:2015tva}, boosted the possible explanation of the Dark Matter problem in terms of Super Heavy Dark Matter, i.e. relic particles created by rapidly varying gravitational fields during inflation (see \cite{Aloisio:2006yi} and reference therein). One of the key expectations of this kind of models is the huge amount of neutrinos (and gamma-rays) produced at energies $\gtrsim10^{20}$ eV \cite{Aloisio:2015lva}. In the present paper we will review recent results that link cosmological observations to the fluxes of neutrinos expected at the highest energies, also assessing the detection capabilities of future UHECR and neutrino observatories.

\section{Secondary cosmogenic neutrinos} 
\label{sec_nu}

The physics of UHECR propagation is well understood \cite{Aloisio:2008pp,Aloisio:2010he,Allard:2011aa}. During their journey from the source to the observer UHECR experience interactions with astrophysical backgrounds\footnote{We will not consider here the interaction with extragalactic magnetic fields, whose presence is not clear and yet under discussion.}, namely the Cosmic Microwave Background (CMB) and the Extragalactic Background Light (EBL). The propagation of UHECR protons\footnote{Hereafter discussing freely propagating UHE nucleons we will always refer only to protons because the decay time of neutrons is much shorter than all other time scales involved \cite{Aloisio:2008pp,Aloisio:2010he,Allard:2011aa}.} is affected almost only by the CMB radiation field and the processes that influence the propagation are: (i) pair production and (ii) photo-pion production \cite{Allard:2011aa,Berezinsky:2002nc}. On the other hand, the propagation of heavier nuclei is affected also by the EBL and the interaction processes relevant are: (i) pair production and (ii) photo-disintegration \cite{Aloisio:2008pp,Aloisio:2010he,Puget:1976nz,Allard:2005ha}. 

The interaction processes that involve UHECRs with background photons are important not only as mechanisms of energy losses affecting the behaviour of the propagating particles but also being the source of secondary particles such as neutrinos, gamma-rays and electron-positron pairs. Here we will focus mainly on the production of secondary neutrinos. The main source of these particles is certainly the process of photo-pion production. A nucleon ($N$), whether free or bounded in a nucleus, with Lorentz factor $\Gamma \gtrsim 10^{10}$ interacting with the CMB photons gives rise to the photo-pion production process: 
\begin{equation}
N+\gamma \to N' + \pi^0   \qquad N+\gamma \to N' + \pi^{\pm}.
\end{equation}
At lower energies $\Gamma< 10^{10}$, even if with a lower probability, the same processes can occur on the EBL field. In the case of UHE protons propagating in the CMB, the photo-pion production process involves a sizeable energy loss producing the so-called GZK cut-off \cite{Greisen:1966jv,Zatsepin:1966jv}, a sharp suppression of the flux of protons expected on Earth at $E\simeq 6\times 10^{19}$~eV. 

The photo-pion production process holds also for nucleons bound within UHE nuclei, being the interacting nucleon ejected from the parent nucleus, but this process is subdominant with respect to nucleus photo-disintegration except at extremely high energies \cite{Allard:2011aa}.

UHE nuclei propagating through astrophysical backgrounds can be stripped of one or more nucleons by the interactions with CMB and EBL photons, giving rise to the process of photo-disintegration:
\begin{equation}
(A,Z) + \gamma \to (A-n, Z-n') + nN
\end{equation}
$n$ ($n'$) being the number of stripped nucleons (protons). In the nucleus rest frame the energy involved in such processes is usually much less than the rest mass of the nucleus itself, therefore in the laboratory frame we can neglect the nucleus recoil. 

Let us now concentrate on the unstable particles (pions, free neutrons and unstable nuclei) produced by the propagation of UHECRs through photo-pion production and photo-disintegration. In most cases the decay length of such particles is much shorter than all other relevant length scales, so these particles decay very soon giving rise to secondary neutrinos.

There are two processes by which neutrinos can be produced in the propagation of UHECRs:
\begin{itemize}
 \item the decay of charged pions produced by photo-pion production, $\pi^{\pm}\to \mu^{\pm} + \nu_\mu(\bar{\nu}_\mu)$, and the subsequent muon decay $\mu^{\pm}\to e^{\pm}+\bar{\nu}_\mu(\nu_\mu)+\nu_e(\bar{\nu}_e)$; 
 \item the beta decay of neutrons and nuclei produced by photo-disintegration: $n \to p + e^{-} + \bar{\nu}_e$, $(A,Z) \to (A,Z-1) + e^{+} + \nu_e$, or $(A,Z) \to (A,Z+1) + e^{-} + \bar{\nu}_e$.
\end{itemize}
These processes produce neutrinos in different energy ranges: in the former the energy of each neutrino is around a few percent of that of the parent nucleon, whereas in the latter it is less than one part per thousand (in the case of neutron decay, larger for certain unstable nuclei). This means that in the interactions with CMB photons, which have a threshold around $\Gamma\gtrsim 10^{10}$, neutrinos are produced with energies of the order of $10^{18}$~eV and $10^{16}$~eV respectively. 

\begin{figure}
\includegraphics[width=0.5\textwidth]{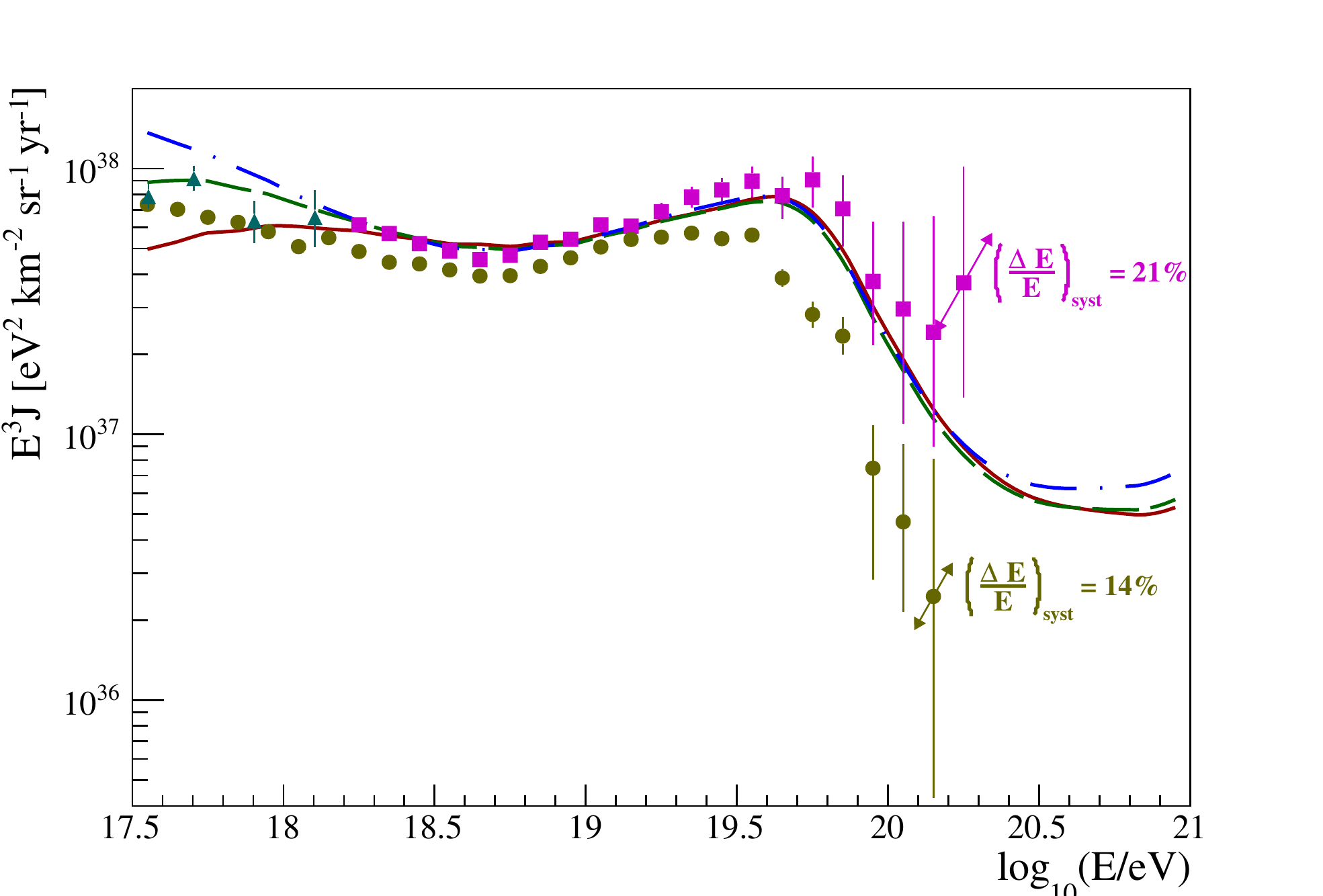} 
\caption{ Fluxes computed in the case of the dip model (pure proton composition) with three different choices for the cosmological evolution of the sources: solid red (no evolution) dashed green (SFR evolution) and dot-dashed blue (AGN evolution). Theoretical fluxes are normalised to TA data (purple filled squares); Auger data (green filled circles) and Kascade-Grande data (blue filled triangles) are also shown for comparison. }
\label{fig1a} 
\end{figure} 

Interactions with EBL photons contribute with a much lower probability than CMB photons, affecting a small fraction of the propagating protons and nuclei. Neutrinos produced through interactions with EBL, characterised by lower thresholds, have energies of the order of $10^{15}$~eV in the case of photo-pion production and $10^{14}$~eV in the case of neutron decay.   

Being weak interacting particles neutrinos, once produced, reach the earth from all over the universe. This is an important point that makes neutrinos a viable probe not only of the chemical composition of UHECRs but also of the cosmological evolution of sources that, as we will show below,  can be also constrained by the neutrino flux observed on Earth. Following \cite{Aloisio:2015ega}, we will consider the case of sources with no cosmological evolution, with the same cosmological evolution as that of active galactic nuclei (AGN), an astrophysical object supposed to play a role in particle acceleration to the highest energies \cite{Berezinsky:2002nc}, and with the cosmological evolution of the star formation rate (SFR).

All computations are performed under the assumption of a homogenous distribution of sources. This assumption does not affect the expected neutrino spectra because in the case of neutrinos the overall universe, up to the maximum redshift, contributes to the fluxes and possible flux variations due to a local inhomogeneity in source distribution gives a negligible contribution to the total flux. We also fix a maximum redshift of the sources $z_{max}=10$, which is the typical redshift of the first stars (pop III) \cite{Berezinsky:2011bb}. In any case the expected fluxes of primary and secondary particles are almost independent of $z_{max}$ if $z_{max}>3$ \cite{Kotera:2010yn,Aloisio:2015ega}. 

Once produced at cosmological distances neutrinos travel toward the observer almost freely, the opacity of the universe to neutrinos being relevant only at the redshifts $z>10$ \cite{Gondolo:1991rn,Weiler:1982qy}. Therefore, given the assumptions discussed above, in our computations we have neglected any effect due to neutrino propagation apart from the adiabatic energy losses due to the expansion of the universe. 

As discussed in the introduction, the experimental evidences about UHECR chemical composition are not conclusive with the two opposite scenarios claimed by TA and Auger. To bracket the expectations in terms of secondary neutrinos we have considered both scenarios of proton-dominated flux (dip model) and mixed composition with heavy nuclei contributing to the flux of UHECR. The results discussed here were already presented elsewhere \cite{Aloisio:2015ega} and obtained using the Monte Carlo code SimProp \cite{Aloisio:2012wj}, properly upgraded to compute the production of secondary neutrinos.

\begin{figure} 
\includegraphics[width=0.5\textwidth]{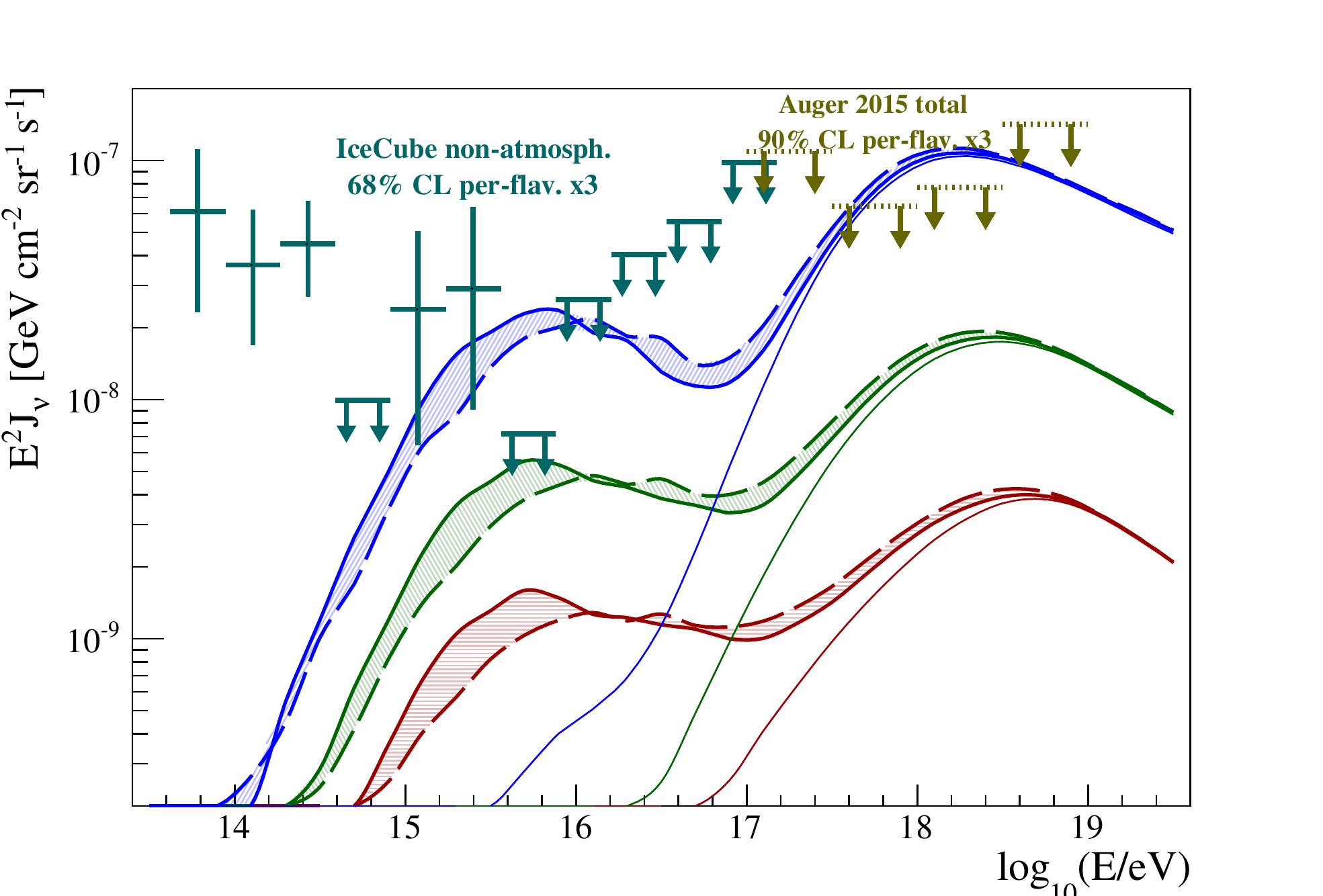}
\caption{Fluxes of secondary neutrinos, corresponding to the fluxes of figure \ref{fig1a} with the same color code, coloured bands show the uncertainties connected with the EBL evolution model considered. Thin solid lines are the neutrino fluxes obtained neglecting the contribution of the EBL radiation field.}
\label{fig1b}
\end{figure}

\subsection{Dip model} 
Retaining the results claimed by TA, under the assumption of pure protons the observations can be very well described in the framework of the dip model \cite{Berezinsky:2002nc,Aloisio:2006wv}, which explains the features observed in the UHECR's spectrum in terms of the sole proton interactions with the CMB background. As discussed above, the effect of EBL on proton propagation has an important role for the production of secondary neutrinos, but it negligibly affects the expected proton flux. 

Figure \ref{fig1a} shows the fluxes of UHECR as measured by TA (purple filled squares), Auger (green filled circles) and Kascade-Grande (blue filled triangles) together with the theoretical fluxes as computed in the framework of the dip model taking the three different cases of sources cosmological evolution as outlined above. In figure \ref{fig1b} we show the fluxes of secondary neutrinos corresponding to the UHECR fluxes of figure \ref{fig1a}, with the same color code, together with the IceCube observations \cite{Aartsen:2013jdh} and the experimental limits on the neutrino flux of Auger \cite{Aab:2015kma}.

As expected a pure proton composition gives a sizeable flux of secondary neutrinos that, at least in the case of strong cosmological evolution, is already at the level of being detected by both IceCube and Auger. This is a non-trivial result that shows how the future detection of UHE neutrinos can be of paramount importance in the physics of UHECR. 

\begin{figure}
\includegraphics[width=0.5\textwidth]{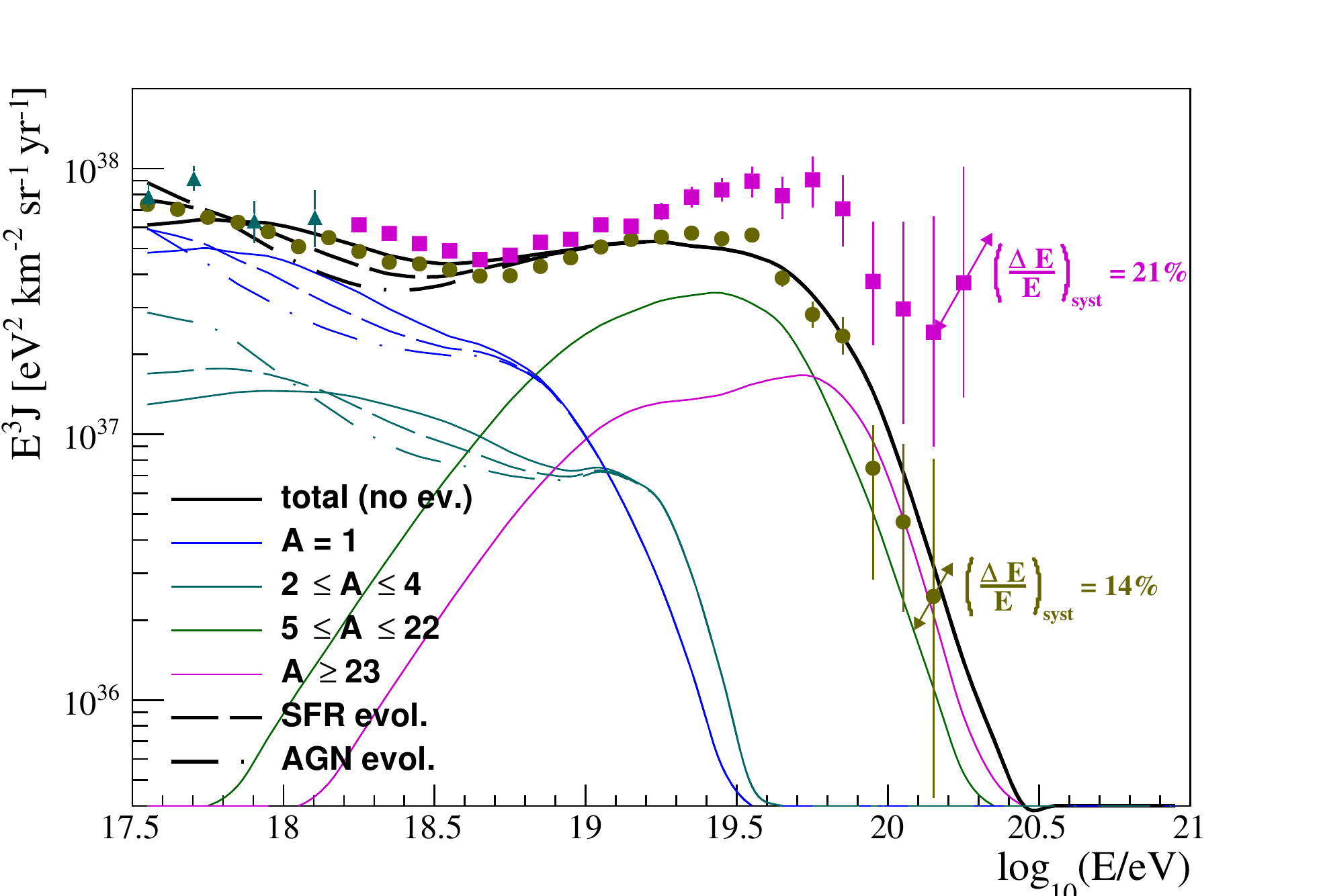} 
\caption{Flux of UHECR in the case of the model with two classes of sources as in \cite{Aloisio:2015ega,Aloisio:2013hya}. Experimental data as in figure \ref{fig1a}.}
\label{fig2a}
\end{figure}

\subsection{Mixed composition}

Changing point of view, taking for granted the results claimed by Auger with heavier composition at increasing energy, the observations can be explained only in terms of a more complicated phenomenology. The ankle feature will be related to a change in the source family contributing to the flux \cite{Aloisio:2015ega,Aloisio:2013hya,Taylor:2011ta} or to specific dynamic at the source \cite{Globus:2015xga,Unger:2015laa}. 

\begin{figure}
\includegraphics[width=0.5\textwidth]{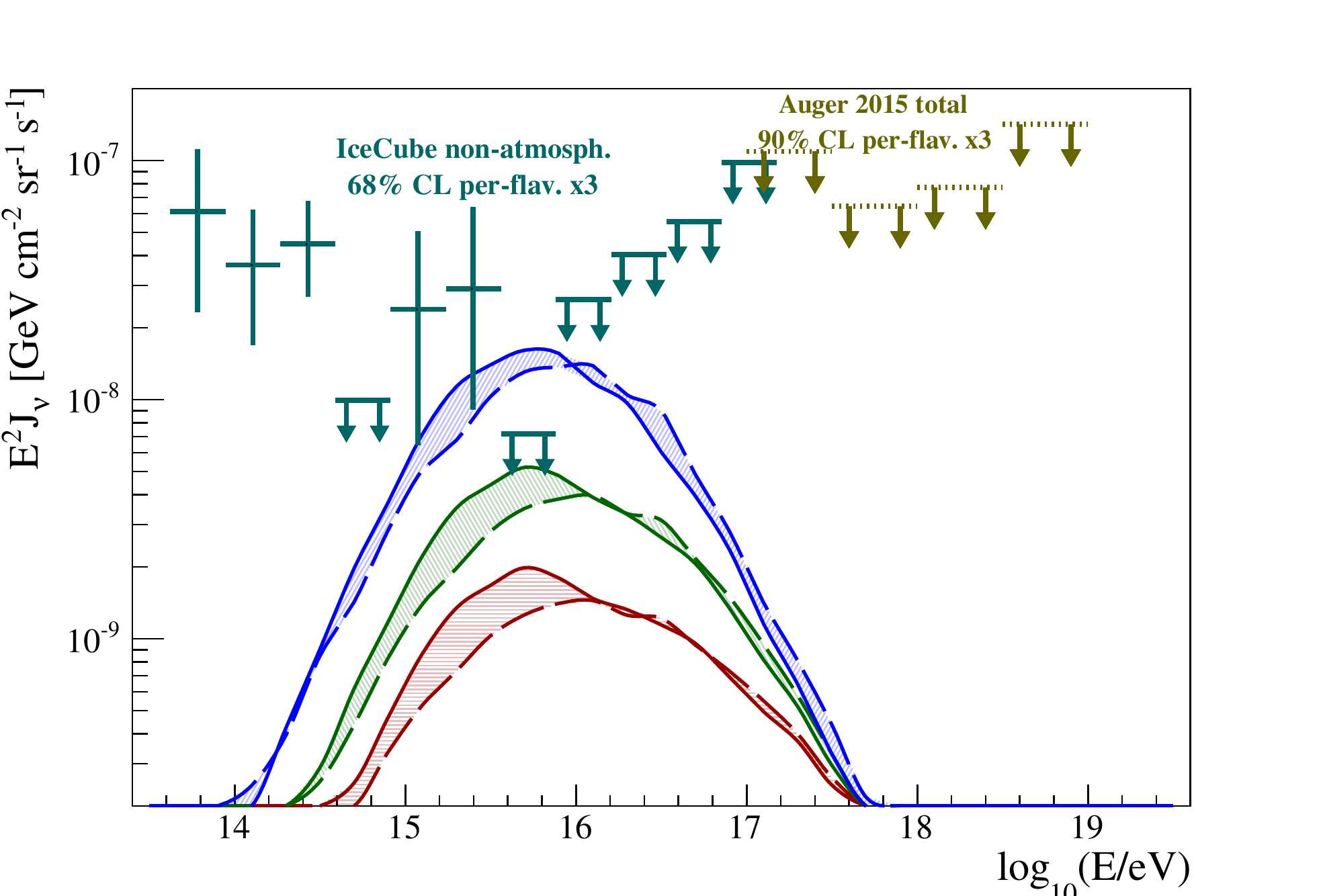}
\caption{ Fluxes of secondary neutrinos, corresponding to the fluxes of figure \ref{fig2a} with the same color code, coloured bands show the uncertainties connected to the EBL evolution model. }
\label{fig2b}
\end{figure}

Taking the model presented in \cite{Aloisio:2015ega,Aloisio:2013hya} in figure \ref{fig2a} we plot the flux of UHECR together with the experimental data of Kascade-Grande, Auger and TA, in figure \ref{fig2b} we plot the corresponding flux of secondary neutrinos. The production of secondary neutrinos in this case is restricted only to PeV energies and comes from the photo-pion production process on the EBL radiation field suffered by protons of the low energy tail of UHECR. In this case the highest energy part of the UHECR spectrum, being dominated by heavy nuclei, does not show any significant contribution to the production of secondary neutrinos.

\section{Super heavy dark matter} 
\label{shdm}

One of the most fundamental and longstanding problems in modern physics is certainly the presence of a yet-not-observed form of matter whose presence is detected only through its gravitational interaction: the so-called Dark Matter (DM) \cite{Bertone:2004pz}. The leading paradigm to explain DM observations is based on the Weakly Interactive Massive Particle (WIMP) hypothesys, which consists of two basic assumptions: (i) WIMPs are stable particles of mass $M_\chi$  that interact weakly with the Standard Model (SM) particles; (ii) WIMPs are thermal relics, i.e. they were in Local Thermal Equilibrium (LTE) in the early Universe. Imposing that the WIMP density today is at the observed DM level, using a simple unitarity argument for the WIMP annihilation cross section $\sigma_{ann} \simeq 1/M_{\chi}^2$, one obtains an estimate of the WIMP mass in the range of $10^2$ to $10^4$ GeV. This result, also called the WIMP ÒmiracleÓ, links the DM problem to the new physics scale expected in the context of the ÒnaturalnessÓ argument for electroweak physics. Triggering in this way the strong hope that the search for WIMP DM may be connected to the discovery of new physics at the TeV scale.

Searches for WIMP particles are ongoing through three different routes: direct detection, indirect detection, and accelerator searches \cite{Bertone:2004pz}. None of these efforts have discovered a clear WIMP candidate so far. In addition, no evidence for new physics has been observed at the Large Hadron Collider (LHC). Although not yet conclusive, the lack of evidence for WIMPs may imply a different solution for the DM problem outside of the WIMP paradigm.

Following \cite{Aloisio:2015lva}, here we will reconsider the scenario based on particle production due to time varying gravitational fields: the so-called Super Heavy Dark Matter (SHDM) scenario. This alternative approach is based on the possibility of particle production due to the non-adiabatic expansion of the background space-time acting on the vacuum quantum fluctuations. It is remarkable the fact that the first attempt to apply such a mechanism in Cosmology dates back to E. Schr\"odinger in 1936 \cite{Schro:1939}. More recently, in the framework of inflationary cosmologies, it was shown that particle creation is a common phenomenon, not tied to any specific cosmological scenario, that can play a crucial role in the solution to the DM problem as SHDM (labeled by X) can have $\Omega_X(t_0)\lesssim 1$ (see \cite{Aloisio:2015lva} and references therein). This conclusion can be drawn under three general hypotheses: (i) SHDM in the early Universe never reaches LTE; (ii) SHDM particles have mass of the order of the inflaton mass, $M_\phi$; and (iii) SHDM particles are long-living particles with a lifetime exceeding the age of the Universe, $\tau_X\gg t_0$. These three hypothesis can be tested experimentally through cosmological and UHECR observations. 

\begin{figure}
\includegraphics[width=0.5\textwidth]{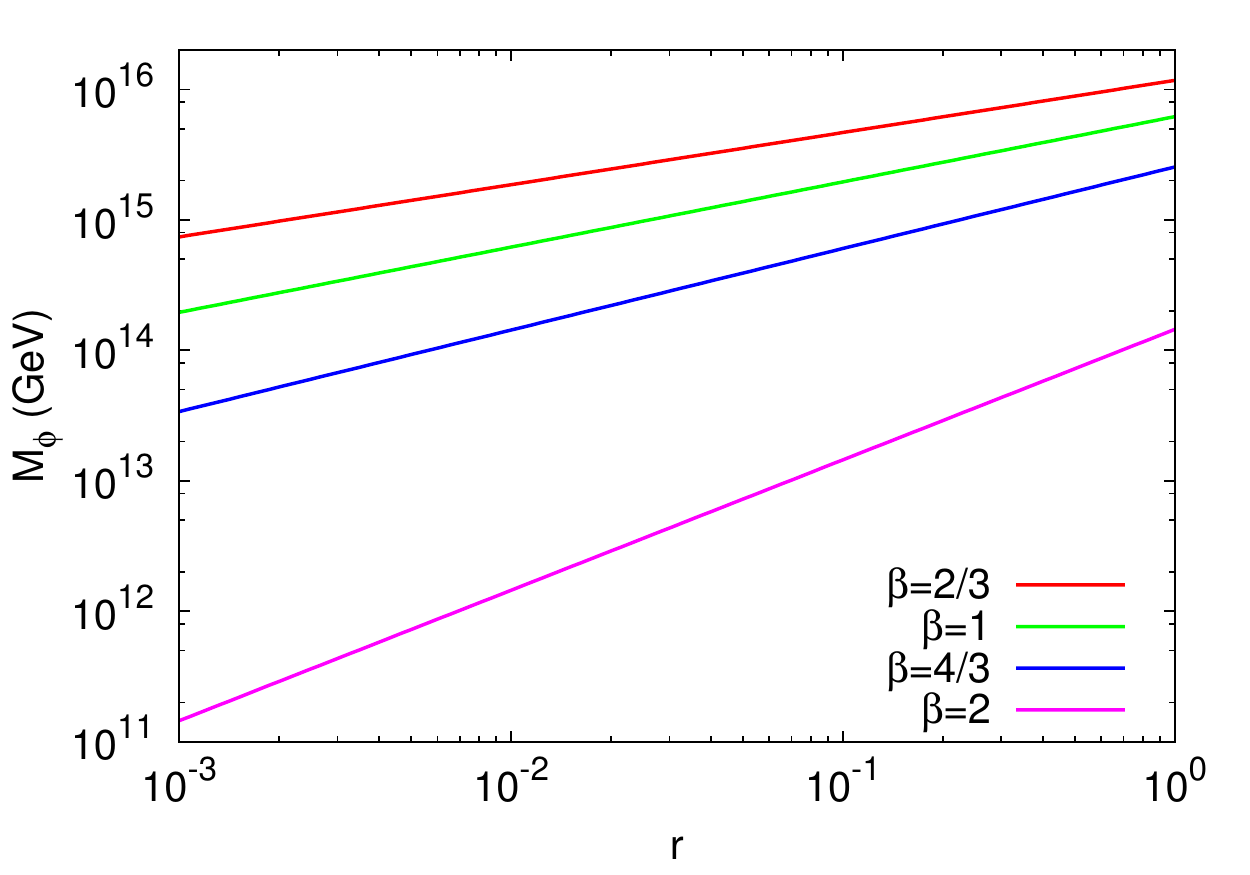} 
\caption{Inflaton mass as function of the ratio $r$ of tensor to scalar modes for different choices of the inflaton potential as labeled.}
\label{fig3a}
\end{figure}

The observations of CMB fluctuations can be directly linked to the primordial density fluctuations in the early universe. These fluctuations are of two types: curvature and iso-curvature. The first being connected with the total energy density in the early universe while the second with the actual composition of the energy density itself. The fact that SHDM particles, with mass $M_X$ of the order of the inflaton mass $M_\phi$, never reach LTE implies a relevant contribution of these particles to the iso-curvature perturbations \cite{Chung:2004nh} that are a source of gravitational potential therefore contributing to the primordial gravitational waves background with a tensor to scalar ratio $r>0$ in the CMB fluctuations. This is an unavoidable consequence of any particle production through time varying gravitational fields. In this sense the recent claim by BICEP2, that claimed a measured value of $r=0.2^{+0.07}_{-0.05}$ \cite{Ade:2014xna}, triggered a renewed interest for the SHDM hypothesis as a viable alternative to the WIMP paradigm. The result claimed by BICEP was reconsidered in the light of a common analysis including Planck and Keck array observations \cite{Ade:2015tva}, validated on simulations of a dust-only modelling and performed by a simple subtraction of scaled spectra. The final result of this combined analysis showed a substantial reduction of the tensor modes in the CMB background with an upper limit $r\le 0.12$, at $95\%$ confidence level, and a likelihood curve that peaks at $r=0.05$ but disfavours zero with a scarce statistical significance.  In the near future several different detectors, both ground-based and sub-orbital, will take data at a variety of frequency enabling a more precise determination of the CMB primordial tensor modes.

\begin{figure} 
\includegraphics[width=0.5\textwidth]{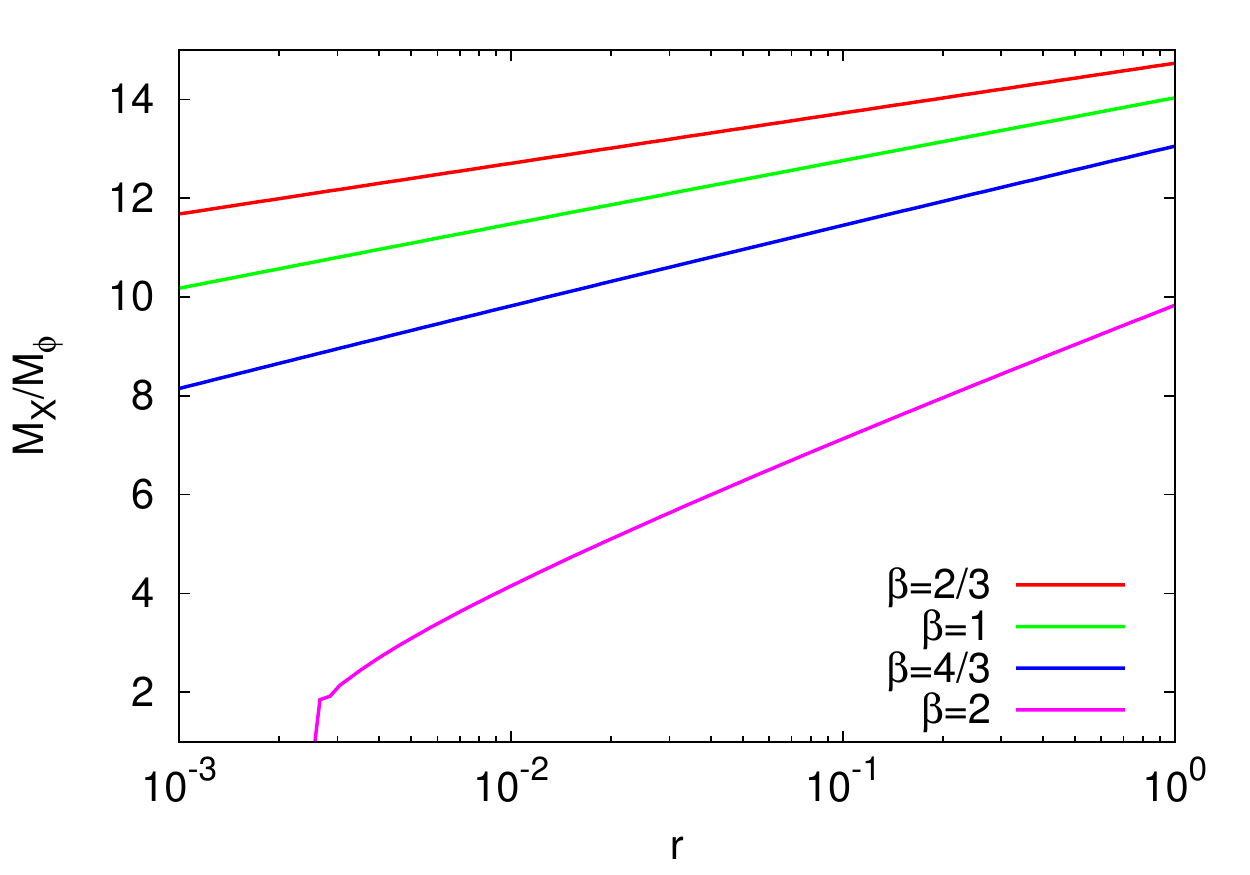}
\caption{ Ratio of the SHDM mass and inflaton mass as function of $r$, obtained as solution of the equation $\Omega_X = \Omega_{DM}$ using different choices of the inflaton potential as labelled.}
\label{fig3b}
\end{figure}

Assuming a non-negligible value of the tensor to scalar ratio $r$ in the CMB fluctuations, one gets the scale of the inflaton mass $M_{\phi}$ and, requiring that the SHDM density today corresponds to the observed DM density $\Omega_X=\Omega_{DM}$, one obtains the scale of the SHDM mass $M_X$. Following the computations of \cite{Aloisio:2015lva}, assuming an inflaton potential of the type $V(\phi)=\phi^\beta M_\phi^{4-\beta}/\beta$ with $\beta=2/3,1,4/3,2$, in figure \ref{fig3a} we show the inflaton mass $M_\phi$ as function of $r$ and in figure \ref{fig3b} the corresponding SHDM mass $M_X$. 

As discussed in \cite{Aloisio:2015lva}, the result presented in figures \ref{fig3a} and \ref{fig3b} also depends on the assumption about the reheating temperature $T_{RH}$ that was assumed to be $T_{RH}=10^{9}$ GeV. For instance, in the case of an inflaton potential $\beta=2$ the equation $\Omega_X=\Omega_{DM}=0.261$ has no solution for $r<3\times 10^{-3}$ with our choice of $T_{RH}=10^{9}$ GeV. In this case, as discussed in \cite{Chung:2004nh}, there is a lower bound on the reheating temperature. Fixing $T_{RH}$ at its minimum the equation $\Omega_X=\Omega_{DM}$ shows no solutions only for a tensor to scalar ratio $r<2\times 10^{-2}$ while usng larger values of $r$ the ratio $M_X/M_\phi$ differs by not more than $50\%$ from the result of figure \ref{fig3b}. In the other cases of $\beta=2/3,1,4/3$ there is no lower bound on $T_{RH}$ and assuming $T_{RH}$ in the range that spans from $10^{5}\div 10^{10}$ GeV the solution of $\Omega_X=\Omega_{DM}$ still differs by not more than $50\%$ from the solution plotted in figure \ref{fig3b}. A change in $M_X$ at the level of a factor 2 does not change the conclusion presented here, therefore we can keep $T_{RH}=10^9$ GeV as a reference value.

\begin{figure} 
\includegraphics[width=0.5\textwidth]{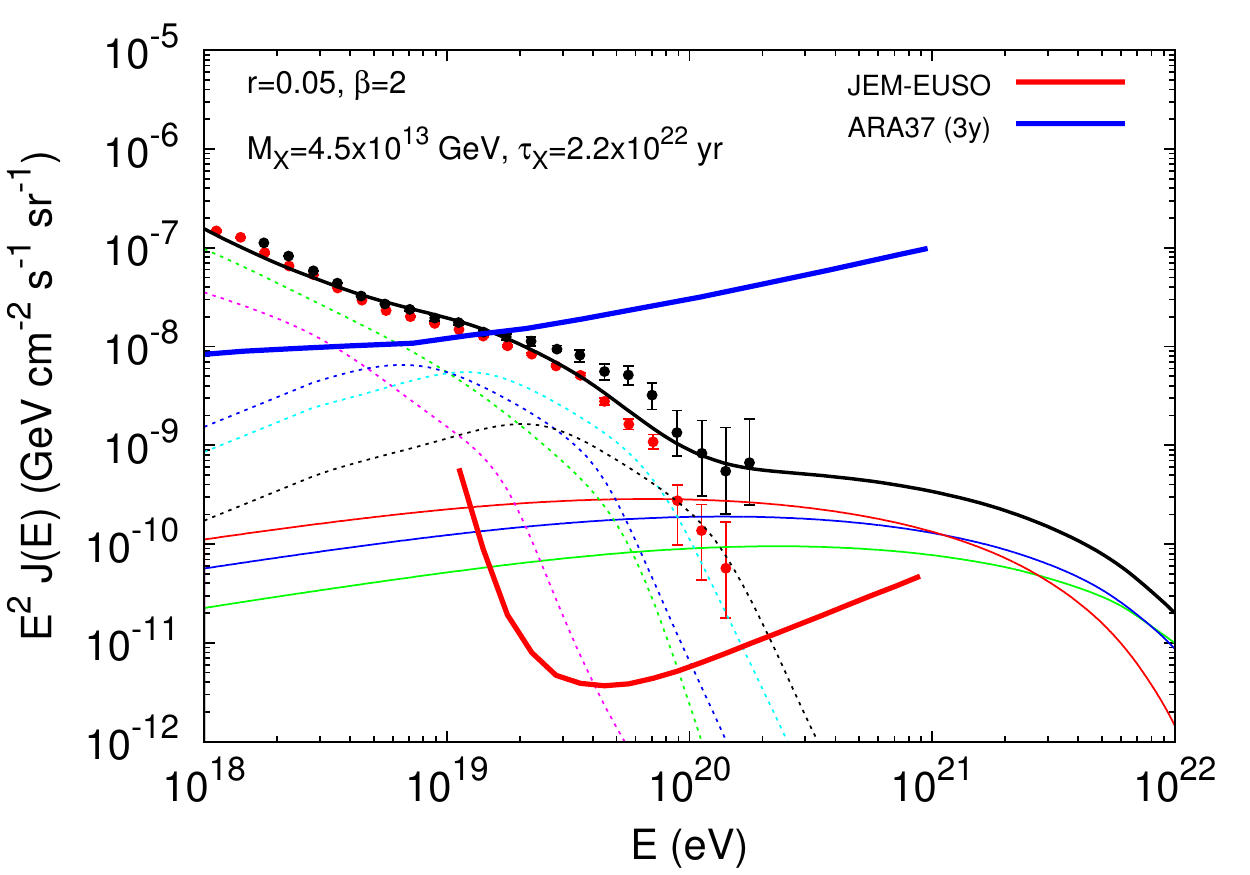}
\caption{ Theoretical fluxes of UHECR from SHDM decay obtained in the case of $r=0.05$ and $\beta=2$, together with UHECR fluxes expected in the framework of the mixed composition model discussed in the previous section. Also shown: Auger data (red points), TA data (black points) and the sensitivity of the future JEM-EUSO space mission (thick red solid line) and, for UHE neutrinos, the upcoming ARA observatory (thick blue solid line).}
\label{fig4a}
\end{figure}

The limits on the SHDM models that come from the experimental limits on the iso-curvature perturbation of the CMB are also satisfied as shown by \cite{Chung:2004nh,Chung:2011xd}. This follows from figure \ref{fig3b} where the ratio $M_X/M_\phi$ needed to obtain the observed DM density today is in the range $2\div 15$ almost independently of the choice of $T_{RH}$ \cite{Aloisio:2015lva}.
 
From figures \ref{fig3a} and \ref{fig3b} we can conclude that SHDM particles production by time-varying gravitational fields provide a viable explanation of the DM problem even with a ratio of tensor to scalar modes at the level of $10^{-3}$. This result still depends on the assumption of a long-living SHDM with a particle life-time $\tau_X$ much longer than the age of the Universe. This last hypothesis can be tested through UHECR experiments by detecting the decay products of SHDM. 

Following \cite{Aloisio:2006yi}, on very general grounds, we can assume that SHDM decay gives rise to a quark anti-quark pair with subsequent parton cascades that, hadronizing, produce Standard Model (SM) particles. The basic signatures of these kind of decays are three: (i) SHDM (as any other DM particle) cluster gravitationally and accumulate in the halo of our Galaxy with an average density $\rho_X^{halo}\simeq 0.3$ GeV/$cm^3$; (ii) in the hadronic cascades the most abundant particles produced are pions, therefore UHE neutrinos and gamma-rays are the most abundant particles expected on Earth, (iii) The non-central position of the Sun in the galactic halo results in an anisotropic flux of the decay products \cite{Aloisio:2007bh}.

\begin{figure} 
\includegraphics[width=0.5\textwidth]{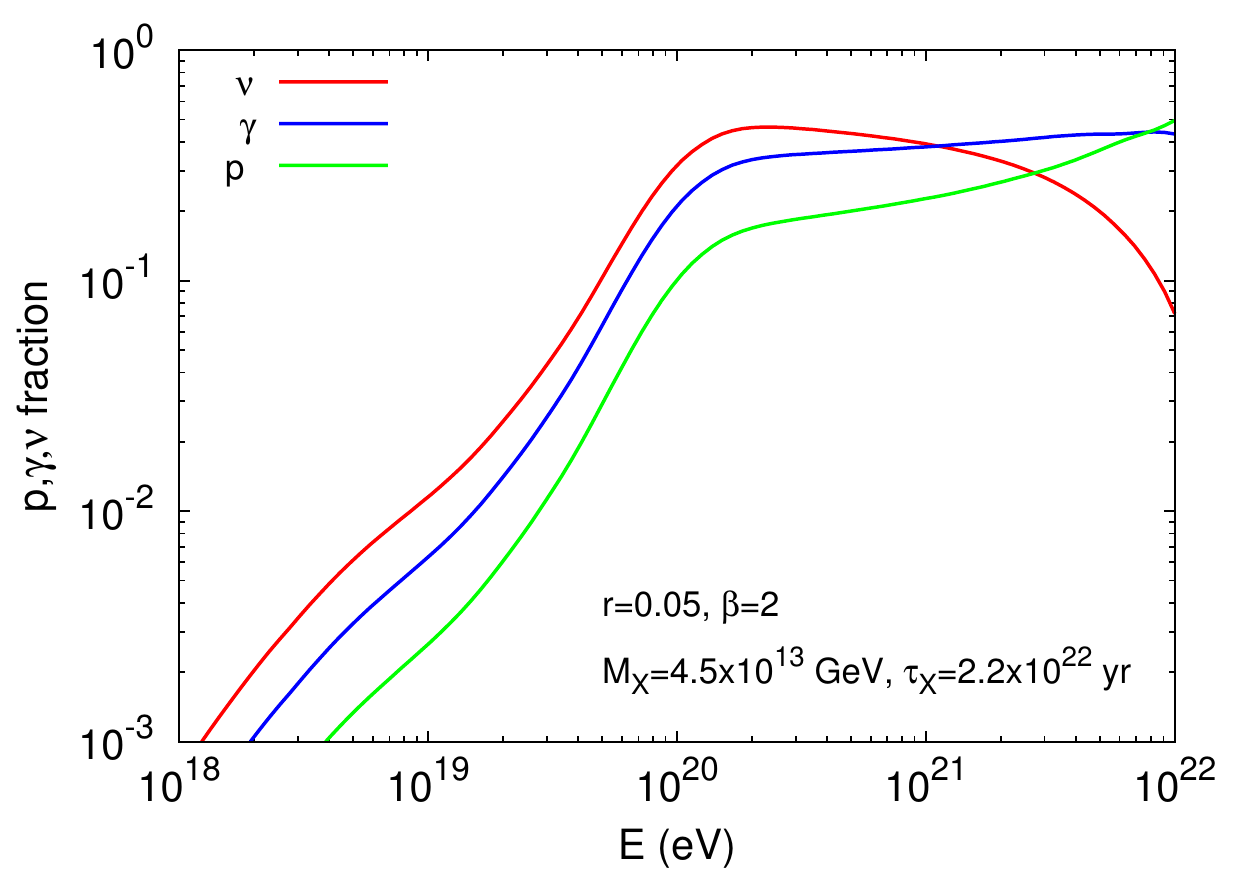}
\caption{ Fraction over the total UHECR flux of protons, photons and neutrinos by SHDM decay, with the same choice of parameters of figure \ref{fig4a}.}
\label{fig4b}
\end{figure}

The quantitative prediction for energy spectra and chemical composition of UHECR pruduced in SHDM decay require an extrapolation of QCD calculations from the TeV scale up to $M_X$. There are several different recipes discussed in literature based on both MC and analytical computations (see \cite{Aloisio:2006yi,Aloisio:2015lva} and references therein) all giving the same behaviour of the expected fluxes $dN/dE\propto E^{-1.9}$, independently of the particle type, with a photon/nucleon fraction $\gamma/N\simeq 2\div 3$ and a neutrino/nucleon fraction $\nu/N\simeq 3\div 4$, quite independent of the energy. 

The UHECR emissivity produced by the decay of SHDM in the halo of our galaxy can be wirtten has 
\begin{equation}
I_{p,\gamma,\nu}(E,R)=\frac{1}{M_X\tau_X}\frac{dN_{p,\gamma,\nu}(E)}{dE}\rho_X(R) 
\label{eqI}
\end{equation}
where $M_X$ and $\tau_X$ are the mass and life-time of the SHDM particle, $dN/dE$ is the energy spectrum of the decay products and $\rho_X(R)$ is the SHDM density in the galactic halo as function of the distance $R$ from the Galactic Center. Here we will assume a DM density profile as comes from numerical simulations by Moore (see \cite{Aloisio:2015lva} and references therein).

As discussed in \cite{Aloisio:2015lva}, the expected flux of UHE particles coming from the SHDM decay can be computed as the integral over the line of sight of the emissivity (\ref{eqI}). In the case of neutrinos, together with particles coming from the local halo, it is also relevant the contribution of the whole universe that is at the level of $10\%$ of the flux coming from our own galaxy \cite{Aloisio:2015lva}. 

Tacking into account the contribution of protons, gammas and neutrinos produced by SHDM decay, in figure \ref{fig4a} we plot the flux of UHECR summing the contribution of the mixed composition model of \cite{Aloisio:2015ega,Aloisio:2013hya} to that coming from SHDM. In figure \ref{fig4a} we also plot the experimental data of Auger (red points) and TA (black points) and the expected sensitivities of ARA (thick solid blu line), a future neutrino observatory based on the Askarian effect \cite{Allison:2011wk}, and JEM-EUSO (thick solid red line), an UHECR detector planned onboard the international space station \cite{Ebisuzaki:2014wka}. The fluxes of UHECR produced by SHDM decay are plotted as solid lines (red for neutrinos, blu for gamma-rays and green from protons), these fluxes are obtained in the case of an inflationary potential $\beta=2$ and a ratio of tensor to scalar modes $r=0.05$ that corresponds to a SHDM mass $M_X=4.5\times 10^{13}$ GeV (as labeled in the plot). The value of $\tau_X=2.2\times 10^{22}$ y in figure \ref{fig4a} is fixed in order not to overshoot the Auger limits on gamma-rays at $\lesssim 10^{19}$ eV.  

From figure \ref{fig4a} follows that future observatories of UHECRs and neutrinos should be able to discover SHDM or constrain its lifetimes. The EUSO detector seems particularly suited for these kind of studies as it achieves about an order of magnitude higher exposure at $10^{20}$ eV respect to Auger. The most striking signature of SHDM models is represented by the peculiar composition of UHECR at the highest energies, with a flux dominated by neutrinos and gamma-rays. This signature is a precise outcome of the sole decaying dynamics of SHDM particles \cite{Aloisio:2006yi} and it is shown in figure \ref{fig4b} where the ratio of neutrinos, gamma-rays and protons coming from SHDM decays is plotted. 

\section{Conclusions}
\label{conc}

The observation of UHE neutrinos, starting from PeV energies up to the highest, is a new window through which we can observe the Universe. This window, recently opened by the IceCube observations of PeV neutrinos, is of paramount importance from many aspects. Here we have reviewed the tight relationship that links UHECR and UHE neutrinos. We have considered two different, possible, sources of UHE neutrinos: namely those coming from the propagation of UHECR and those coming from the decay of SHDM. 

Neutrinos coming from the propagation of UHECR originate mainly by the decay of charged pions produced by photo-hadronic interactions of protons and heavy nuclei with CMB and EBL backgrounds. As sources of neutrinos, these processes are efficient only in the case of protons while in the case of heavier nuclei photo-hadronic interactions are significantly suppressed. This is an important outcome that signals the strong link between UHECR chemical composition and the observation of UHE neutrinos. 

The light composition of the low energy tail of UHECR is observed with a solid consensus by all detectors. Therefore a flux of UHE (cosmogenic) neutrinos in the PeV energy range is assured, these particles come from the photo-pion production process suffered by EeV protons on the EBL photons. The remaining ignorance in predicting the flux of neutrinos is connected with the cosmological evolution of sources and, to a lesser extent, with the uncertainties in the cosmological evolution of the EBL background itself. Models with a strong cosmological evolution of sources, as in the case of AGN, produce a neutrino flux almost at the level of the IceCube observations in the PeV region. This fact already enables a partial constraining of cosmological evolution of sources, disfavouring models with too strong evolution. 

At the highest (EeV) energies the flux of cosmogenic neutrinos mainly originates from the highest energy tail of the UHECR spectrum. If the most energetic cosmic rays, at around $10^{20}$ eV, are mainly protons, then a sizeable flux of EeV neutrinos will be produced. In this case, as before, the experimental limits on EeV neutrinos can be used to constrain the cosmological evolution of sources. For instance, Auger limits on neutrino fluxes already disfavour too strong evolution models and moderate cosmological evolution can be tested by future UHE neutrino detectors. Conversely, if the highest energy tail of UHECRs is composed mainly of heavy nuclei, as in models reproducing Auger data on spectrum and chemical composition, then the flux of cosmogenic EeV neutrinos is far below the detection threshold of any running or planned detector.

The observation of UHE neutrinos is also important to test new physics models, as those connected with SHDM and Cosmology. The BICEP program revived the notion that the primordial B-mode polarization may be higher than previously expected. However, Planck showed that subtracting foreground contamination by dust is also more challenging than past estimates. The joint Planck, BICEP2, and Keck Array analysis set an upper limit on $r$ while mildly pointing towards a non-zero value. 

Given the large interest in the community and the ability of next generation experiments to surpass current challenges, it is likely that the fundamental measurement of $r$ will be reached in the near future.
Relatively large values of r, at the level $r>10^{-3}$, motivate the idea that dark matter is mainly composed of superheavy relics from the inflationary epoch. Given a measurement of $r$, the existence of SHDM can best be tested by exploring the decay lifetime parameter range with future UHECR observatories. The higher the statistics of UHECR experiments at energies around $10^{20}$ eV, the more likely the detection of SM particles produced by SHDM decay. Given the currently planned UHECR and UHE neutrino detectors, JEM-EUSO is best placed to effectively study the allowed SHDM lifetimes with possible lifetime detections or constraints reaching values as high as $\tau_X \simeq 10^{24}$ y.

\section*{Aknowledgements}
It is a pleasure to thank the co-authors of the results discussed here, namely: D. Boncioli, A. di Matteo, A. Grillo, S. Matarrese,  A. Olinto and S. Petrera. I'm also grateful to V. Berezinsky and P. Blasi for continuos collaboration in the field of UHECR. 

\bibliographystyle{elsarticle-num}
\bibliography{UHECR}
\end{document}